\documentclass[prl,aps]{revtex4}
\begin{document}
\title{Violations of conservation laws in viscous liquid dynamics}
\author{Jeppe C. Dyre}
\affiliation{DNRF centre  ``Glass and Time'', Building 27 (IMFUFA), 
Roskilde University, Postbox 260, DK-4000 Roskilde, Denmark}
\date{\today}
\newcommand{\xik}{\xi_{\bf k}}
\newcommand{\iomt}{i\omega\tau}
\newcommand{\bu}{{\bf u}}
\newcommand{\br}{{\bf r}}
\newcommand{\bnul}{{\bf 0}}
\newcommand{\bk}{{\bf k}}
\newcommand{\bom}{{\bf\Omega}}
\newcommand{\bnabla}{{\bf\nabla}}
\newcommand{\thn}{\theta_0}
\newcommand{\la}{\left\langle}
\newcommand{\ra}{\right\rangle}
\newcommand{\rk}[1]{{\rho_{\bf  {#1}}}}
\newcommand{\fik}[1]{{\Phi_{\bf  {#1}}}(t)}
\newcommand{\fnik}[1]{{\Phi^{(0)}_{\bf  {#1}}}(t)}
\newcommand{\rok}{\rk\bk}
\newcommand{\romk}{\rk{-\bk}}
\newcommand{\gr}{\Gamma_{\rho}}
\newcommand{\sti}{\tilde s}
\newcommand{\tti}{\tilde t}
\newcommand{\mt}{\langle\rho\rangle}
\newcommand{\fnul}{\Phi}
\begin{abstract}
The laws expressing conservation of momentum and energy apply to any isolated system, but these laws are violated for highly viscous liquids under laboratory conditions because of the unavoidable interactions with the measuring equipment over the long times needed to study the dynamics. Although particle number conservation applies strictly for any liquid, the solidity of viscous liquids implies that even this conservation law is apparently violated in coarse-grained descriptions of density fluctuations. 
\end{abstract}
\maketitle

Glasses are made by cooling viscous liquids until they solidify \cite{kau48,har76,bra85,ell90,ang95,deb96,ang00,alb01}. Any liquid may form a glass if it is cooled rapidly enough to avoid crystallization \cite{tam25}. In order to understand the properties of glasses one must understand the highly viscous liquid phase preceding glass formation. Viscous liquids and the glass transition continue to attract a great deal of interest because there is still no generally accepted description of viscous liquid dynamics, i.e., their equilibrium fluctuations and corresponding linear response properties. The question we address here is: On which principles should a realistic description of viscous liquids be based? ``Ordinary'' less viscous liquids like ambient water are fairly well understood; they are described by equations based on the fundamental laws expressing conservation of momentum, energy, and particle number \cite{boonyip,hanmc,barrat}. We argue below that viscous liquids are qualitatively different, and that the basic conservation laws cannot be used for arriving at a proper description of viscous liquid dynamics. 

As the glass transition is approached the liquid viscosity increases enormously. Highly viscous liquids exhibit universal features as regards their physical properties. Thus whether involving covalent, ionic, metallic, van der Waals, or hydrogen bonds, virtually all viscous liquids share the {\it three non's}: a) {\it Non}-exponential time dependence of relaxations; b) {\it Non}-Arrhenius temperature dependence of the viscosity with an activation energy that increases as temperature decreases (this applies for the vast majority of viscous liquids -- exceptions are, e.g., pure silica which is virtually Arrhenius); c) {\it Non}-linearity of relaxations following even relatively small temperature jumps. The last point is perhaps least puzzling given the fact that this non-linearity in most models follows directly from the dramatic temperature dependence of the viscosity, but the first two {\it non}'s are crucial characteristics of viscous liquids.

The most sophisticated and theoretically well-based theory for viscous liquid dynamics is the mode-coupling theory \cite{got92,das04}. After a series of approximations starting from Newton's second law, mode-coupling theory ends up with a self-consistent, nonlinear equation for the time dependence of the density autocorrelation function. The strength of the theory is evidenced by the fact that it uniquely predicts the dynamics from the static structure factor and a few other static equilibrium averages. Comparisons to computer simulations and to experiments on moderately viscous liquids generally favour 
mode-coupling theory. Nevertheless, it seems that mode-coupling theory is unable to predict the properties of the highly viscous liquid phase preceding conventional glass formation. As argued below, mode-coupling theory's starting point of the momentum, energy and particle conservation laws do not apply for highly viscous liquids under realistic conditions. This may explain why the theory apparently only works for the less-viscous phase.

A liquid is characterized by several diffusion constants. Of primary interest here are the single-particle diffusion constant $D_s$, which determines the mean-square displacement at long times via $\langle \Delta x^2(t)\rangle=2D_st$, and the transverse momentum diffusion constant: the kinematic viscosity $\nu$ of the Navier-Stokes equation. For ``ordinary'' liquids like ambient water -- with viscosity in the $10^{-3}$ Pa s range -- these diffusion constants typically have values within one or two orders of magnitude of $10^{-7}$ ${\rm m^2/s}$. This is easily understood from elementary kinetic theory, according to which the diffusion constant is of order the mean-free path squared divided by the mean time between collisions of the diffusing entity. Rough estimates of these quantities are 1 Angstrom and 0.1  picosecond respectively, resulting in the value $10^{-7}$ ${\rm m^2/s}$.

For liquids approaching the glass transition the diffusion constants decouple. The single-particle diffusion constant decreases, whereas the kinematic diffusion constant increases. Just above the glass transition the viscosity is typically of order $10^{12}$ Pa s, which is a factor $10^{15}$ larger than for ``ordinary'' liquids. Thus if the single-particle diffusion constant is taken to be inversely proportional to the viscosity as predicted by the Debye-Stokes-Einstein equation \cite{dse}, the ratio $\lambda=D_s/\nu$ decreases from roughly $1$ to roughly $10^{-30}$ close to the glass transition. Such small dimensionless numbers are highly unusual in condensed matter physics. The small magnitude of $\lambda$ strongly suggests that there is a qualitative change of behaviour going from ``ordinary'' to extremely viscous liquids.

The question is how this qualitative change manifests itself. The Navier-Stokes equation most likely still applies for extremely viscous liquids, although this remains to be proven experimentally. The extremely high viscosity, however, itself hints at new and different physics of viscous liquids compared to ``ordinary'' liquids: As an example, if the viscosity is $10^{10}$ Pa s and the density is $10^3$ kg/${\rm m^3}$, the dynamic viscosity is $10^7$ ${\rm m^2/s}$. Typical alpha relaxation times for this viscosity are of order seconds, which is thus the relevant time scale for studying viscous liquid dynamics. In this case transverse momentum diffuses more than one kilometre over the alpha relaxation time, so for typical laboratory sample sizes momentum has ample time to diffuse via the sample holder into and out of the liquid. This means that liquid momentum cannot be regarded as conserved and that consequently descriptions based on momentum conservation are misleading \cite{III}.

To make these considerations more precise, note first that for any physical quantity $A$ with zero mean, if square brackets denote an equilibrium average, at any given temperature we can define a characteristic time $\tau_A$ by

\begin{equation}\label{1}
\tau_A\,=\,
\frac{1}{\langle A^2\rangle}\int_0^\infty \langle A(0)A(t)\rangle dt\,.
\end{equation}
This definition applies, e.g., for the liquid's total electrical dipole moment, its average shear stress, its volume, etc, but it also applies for the liquid's momentum and energy (all quantities are understood to have been  subtracted their mean value). The momentum characteristic time, $\tau_{\rm mom}$, is sample-size dependent, of course, varying as for any diffusion process with sample dimension $L$ as $\propto L^2$. For an ``ordinary'' liquid like ambient water, $\tau_{\rm mom}$ is much larger than the dipole characteristic time, which is in the picosecond range. Thus for the calculating the frequency-dependent dielectric constant via the Kubo formula involving the dipole autocorrelation function, because it almost doesn't change over the physically relevant time, momentum may be regarded as constant and conserved. The situation is the opposite for viscous liquids. Taking the above example of a liquid with alpha relaxation time of one second, the momentum characteristic time is here much {\it smaller} than the dipole characteristic time. Consequently, momentum may not be regarded as conserved over the time span relevant for calculating the frequency-dependent dielectric constant. In other words, just as there are adiabatic and isothermal static linear responses, there are also two kinds of 
frequency-dependent linear response functions: Those that refer to a system with constant momentum (``adiabatic'' type linear response functions), and those that refer to a system which continuously exchanges momentum with its surroundings. The latter are relevant for viscous liquids, but few other places. It is possible that these two response functions turn out to be identical in some cases, but it is not obvious that this should be so -- whereas it obviously {\it is} the case for less-viscous liquids.

The next variable to consider is energy. Heat conduction is notoriously slow and the heat diffusion constant is almost independent of the viscosity. Nevertheless, because a typical layer thickness in, e.g., dielectric relaxation measurements is 100 $\mu$m or smaller, when the alpha relaxation time becomes larger than one second, heat exchange to and from the measuring cell becomes unavoidable. Thus eventually, when the viscosity becomes sufficiently large, $\tau_{\rm en}$ becomes smaller than the liquid's alpha relaxation time, and the situation is analogous to that of momentum non-conservation: Energy conservation is violated for viscous liquids under laboratory conditions, and descriptions based on energy conservation may lead to erroneous results.

We proceed to discuss density fluctuations. Molecules in an ``ordinary'' liquid basically move by a superposition of Brownian and vibrational motions. In contrast, molecules in viscous liquids effectively move much more slowly because almost all motion goes into vibrations. These vibrations take place around the ``inherent states'' introduced by Stillinger and Weber defined as the potential energy minima and their basins of attraction \cite{sti83}. Only rarely does anything happen in the form of a so-called flow event, a sudden rearrangement of molecules. Flow events are rare simply because the energy barriers to be overcome are large compared to $k_BT$. This picture of viscous liquids goes back to Kauzmann, who in 1948 described flow events as ``jumps of molecular units of flow between different positions of equilibrium in the liquid's quasicrystalline lattice'' \cite{kau48}. Goldstein in 1969 first emphasized the importance of potential energy minima in configuration space and expressed the belief that ``the existence of potential energy barriers large compared to thermal energy are intrinsic to the occurrence of the glassy state, and dominate flow, at least at low temperatures'' \cite{gol69}. Since then extensive computer simulations have clearly confirmed this picture \cite{heu97,sas98,ang00a,sch00} (although it is still not possible to simulate the very high viscosities reached when the calorimetric glass transition is approached).

In between flow events a viscous liquid is indistinguishable from an amorphous solid; in fact any attempt to simulate a viscous liquid with viscosity close to the laboratory glass transition on present-day computers would show nothing but vibrating molecules. Based on this, it appears that a viscous liquid is more like a solid than like ordinary less-viscous liquids -- albeit a solid that flows \cite{note}:

\[
{\bf Viscous\,\, liquid\,\,\cong\,\,Solid\,\, that\,\, flows}\,.
\]
The ``solidity'' of viscous liquids \cite{I} has an important consequence for the description of density fluctuations: The small molecular displacements in the surroundings of a flow event may be estimated by reference to solid-state elasticity theory. To leading order in $r^{-1}$ the molecular displacement following a localized disturbance in an elastic solid varies as $r^{-2}$ where $r$ in the present case is the distance to the flow event \cite{III,lan70}. More precisely, if the flow event takes place at the orgin and the displacement vector is $\bu$, the leading order term is $\bu\propto\br/r^3$.  Like the Coulomb electrical field $\bu(\br)$ has zero divergence, implying that to leading order there are no long-ranged density changes induced by a flow event. 

In highly viscous liquids flow events may be regarded as instantaneous on the time scale of the alpha relaxation process below the solidity length (introduced in Ref. \cite{I}). Numbering the flow events consecutively after the time they take place, $t_\mu$, if $\br_\mu$ is the centre of the $\mu$'th flow event and $b_\mu$ measures its intensity, the above considerations translate into the following expression for the time derivative of the density $\rho(\br,t)$ in a coarse-grained description:

\begin{equation}\label{2}
\dot\rho(\br,t)\,=\,\sum_\mu b_\mu\delta(\br-\br_\mu)\delta(t-t_\mu)\,.
\end{equation}
This equation does not constitute a dynamic theory, because it contains no information about the correlations between different flow events. Nevertheless, it serves to emphasize that density in viscous liquids has the appearance of a non-conserved field variable: Density can change at one point in space while not changing at any other point; thus if the individual molecular motions are not traced but only the coarse-grained density profile, it would appear that molecules have been created or annihilated. Another way of seeing that Eq. (\ref{2}) corresponds to a non-conserved variable is that when it is transformed into k-space, this equation corresponds to decay rates that are k-independent \cite{III} whereas conserved variables have decay rates proportional to $k^2$.

In conclusion, viscous liquids are qualitatively different from the less-viscous liquids studied in conventional liquid state theory. A proper description of viscous liquid dynamics should take into account the fact that the fundamental conservation laws do not hold under laboratory conditions. Thus it appears that Newton's laws do not provide a useful starting point whereas, e.g., a stochastic description is more appropriate. The obvious choice here are Langevin equations for the relevant variables (density, dipole density, stress tensor, etc) \cite{III}, or the equivalent description of the probability functional in terms of the corresponding Smoluchowski equations.

\acknowledgments 
This work was supported by a grant from the Danish National Research Foundation (DNRF) for funding the centre for viscous liquid dynamics ``Glass and Time.''

\vspace{1.5cm}

\framebox[\width][c]{
\begin{tabular}{lll}
{\phantom{}} & {\hspace{1cm}\bf ``Ordinary'' liquids} & {\hspace{1cm}\bf Viscous liquids}\\
\hline
{\it Relaxation time} & {\hspace{1cm}Picoseconds} & {\hspace{1cm}Seconds, years, ...}\\
{\it Momentum conservation} & {\hspace{1cm}Crucially important} & {\hspace{1cm}Disobeyed}\\
{\it Energy conservation}& {\hspace{1cm}Crucially important} & {\hspace{1cm}Disobeyed}\\
{\it Particle number conservation}& {\hspace{1cm}Crucially important} & {\hspace{1cm}Apparently disobeyed}\\
\end{tabular}
}

\vspace{1.5cm}

{\small{\bf Fig. 1}$\,$ Basic conservation laws are violated in viscous liquids under realistic laboratory conditions: Both momentum and energy are unavoidably exchanged with the measuring equipment over the relevant time scale (the alpha relaxation time). Particle number is strictly conserved, of course, but nevertheless the solidity of viscous liquids implies that density in a coarse-grained description has the appearance of  non-conserved field.}

\vspace{1cm}


\begin{thebibliography}{99}
\bibitem{note} Phenomenologically one often defines a solid as a system which does not flow, implying that a ``solid that flows'' is a contradiction in terms. Whether a system flows or not, however, is a matter of time scale; even a crystal flows at finite temperature: In a gravitational field, for instance, a vertically flat crystalline sample has lower free energy than, e.g., a cubic-shaped sample. Since any crystal in thermal equilibrium contains point defects, and since their movements are impeded only by finite energy barriers, ``flow'' from the cubic-shaped sample to the flat is not only possible but bound to take place over sufficiently long time scales.
\end{thebibliography}
\end{document}